\documentclass[10pt,pra,twocolumn,superscriptaddress]{revtex4}

\usepackage{graphicx}
\usepackage{amssymb}
\usepackage{times}
\usepackage{amsmath}

\begin{document}
    
\title{Continuous Variable Quantum State Sharing via Quantum Disentanglement}

\author{Andrew M. Lance} \affiliation{Quantum Optics Group, Department
of Physics, Faculty of Science, Australian National University,
ACT 0200, Australia}

\author{Thomas Symul} \affiliation{Quantum Optics Group, Department of
Physics, Faculty of Science, Australian National University, ACT
0200, Australia}

\author{Warwick P. Bowen} \affiliation{Quantum Optics Group,
Department of Physics, Faculty of Science, Australian National
University, ACT 0200, Australia} 
\affiliation{Quantum Optics Group, Norman Bridge Laboratory of Physics
California Institute of Technology, Pasadena, CA 91125, U.S.A.} 

\author{Barry C. Sanders} \affiliation{Institute for Quantum Information Science,
University of Calgary, Alberta T2N 1N4}

\author{Tom\'a\v{s} Tyc} \affiliation{Institute of Theoretical Physics, 
Masaryk University, 61137 Brno, Czech Republic}

\author{T. C. Ralph} \affiliation{Department of Physics, University of Queensland, St Lucia QLD 4072, Australia}

\author{Ping Koy Lam} \affiliation{Quantum Optics Group, Department of
Physics, Faculty of Science, Australian National University, ACT
0200, Australia}

\begin{abstract}
Quantum state sharing is a protocol where perfect reconstruction of quantum states is achieved with incomplete or partial information in a multi-partite quantum network.  Quantum state sharing allows for secure communication in a quantum network where partial information is lost or acquired by malicious parties.  This protocol utilizes entanglement for the secret state distribution, and a class of ``quantum disentangling" protocols for the state reconstruction.  We demonstrate a quantum state sharing protocol in which a tripartite entangled state is used to encode and distribute a secret state to three players. Any two of these players can collaborate to reconstruct the secret state, whilst individual players obtain no information.  We investigate a number of quantum disentangling processes and experimentally demonstrate quantum state reconstruction using two of these protocols.  We experimentally measure a fidelity, averaged over all reconstruction permutations, of $\mathcal {F}= 0.73\pm0.02$.  A result achievable only by using quantum resources.

\end{abstract}

\date{\today} \maketitle

\section{Introduction}

The advent of quantum information science has heralded 
the birth of two exciting new fields of research in quantum mechanics: 
{\it quantum computation} and {\it quantum information networks} \cite{Nie00}. 
Quantum computation involves computation via quantum 
mechanical techniques, using quantum states known as qubits, to 
outperform conventional computers for certain computational problems~\cite{Sho94,Gro97}. 
Quantum information networks, the quantum analogy of the 
internet, are expected to  consist of nodes, where information is 
processed and stored, connected by quantum channels, through which 
quantum information can be transmitted. 
Both quantum computation and quantum information networks
share several key similarities, as they are both concerned with the 
creation, processing and distribution of quantum states. 
They are, however,  both
vulnerable to the loss or destruction of quantum states: through 
de-coherence, node or channel failures, or the intervention 
of malicious parties. 
For this reason protocols that allow for the secure and 
robust distribution of quantum states are vital for the successful 
implementation of these protocols. 

In computer science, Shamir~\cite{Sha79} proposed {\it secret sharing} as a 
protocol that enables the secure distribution of
classical information in networks. Secret sharing can be used to enhance
the security of communication networks such as the internet, 
telecommunication systems and distributed computers.
Quantum resources allow the extension of secret sharing
into the quantum domain in one of two ways. The first involves 
using quantum resources to enhance the security of classical 
information in crypto-communication systems,  
and is known as {\it quantum secret sharing}~\cite{Hil99,Kar99,Tit01}. 
The second uses
quantum resources to securely 
encode and distribute 
quantum states. This second class,   
which we term {\it quantum state sharing} to distinguish from the 
first class of protocols,
is of more significance to 
quantum information protocols, 
which are primarily concerned with 
quantum states. 
In $(k,n)$ threshold quantum state sharing, 
originally proposed by 
Cleve {\it et al.}~\cite{Cle99}, a secret
state is encoded by  the ``dealer'' 
into an $n$-party entangled state or  ``share".
Any $k$ players (the authorized group) can collaborate to 
retrieve the quantum state, whereas the 
remaining $n-k$ players (the adversary group), even when 
conspiring, acquire nothing.  
As a consequence of the no-cloning theorem, 
the number of players in the authorized group 
must consist of a majority of the players, $(k>n/2)$.      
For quantum computation and quantum information networks, quantum 
state sharing provides a secure framework 
for distributed quantum communication, protecting the quantum 
states from the loss up to $n\!-\!k$ shares due to destruction, failures, or malicious conspiracies. 

In general, most theoretical proposals for quantum 
state sharing, by Cleve {\it et al.}~\cite{Cle99}
and other subsequent theoretical proposals~\cite{Sin04, Ima03, Hay04, Man04}, 
are formulated
for the discrete regime. These proposals require 
qudits (multi-dimensional qubits) for the encoding and distribution of
the secret quantum states. 
Experimentally, however, the control and coupling of qudits 
is extremely challenging, making an experimental demonstration of
quantum state sharing in the discrete regime a difficult task. 
Recently, Tyc and Sanders~\cite{Tyc02} extended quantum state sharing to the 
continuous variable regime. Their proposal utilizes 
Einstein-Podolsky-Rosen (EPR) entanglement, an experimentally 
accessible quantum resource~\cite{Ou92, Bow03}, which has been used in several
quantum information protocols including quantum teleportation~\cite{Fur98}, quantum
dense coding~\cite{Li02} and entanglement swapping~\cite{Jia04}.  
Importantly, Tyc {\it et al.}~\cite{Tyc03}, later showed that continuous variable quantum
state sharing could be extended to a $(k,n)$ threshold scheme,
without a corresponding scale up in quantum resources. 
This makes quantum state sharing an important and powerful security protocol 
for future quantum information systems.

In this paper we experimentally demonstrate $(2,3)$ threshold quantum state 
sharing in the continuous variable regime~\cite{Lan04}. In our scheme, a 
secret coherent state is encoded into a tripartite 
entangled state and distributed to three players. 
In general, arbitrary quantum states can be shared via quantum state sharing.
Experimentally we demonstrate quantum state sharing using secret coherent states with unknown coherent amplitude and phase displacements. The coherent states form an over-complete basis, making it possible to infer performance for arbitrary input states from our results.    
We demonstrate that any two of the three players can form an authorized group 
to reconstruct the state, and characterize this state reconstruction in terms of 
fidelity ($\mathcal{F}$), signal transfer ($\mathcal{T}$), and reconstruction noise ($\mathcal{V}$).  
These measures show a direct verification of our tripartite continuous 
variable entanglement. 
The entangled state in the dealer protocol ensures that the 
quantum features of the secret state can be reconstructed by the authorized group,
whilst simultaneously providing security against individual players.
We also demonstrate that security of our scheme 
can be enhanced using classical encoding techniques. 

This paper is presented in the following manner, in Section II we describe the dealer protocol for encoding and distributing the secret state to the players, and we describe a set of ``disentangling protocols" that can be used to reconstruct the secret state by the corresponding authorized groups. In Section III 
we present techniques to characterize the 
state reconstruction. In Sections IV and V we describe the experimental setup and present the experimental results. Finally we conclude in Section VI.
\section{(2,3) Quantum State Sharing Protocols}
\begin{figure}[ht]
\includegraphics[width=8cm]{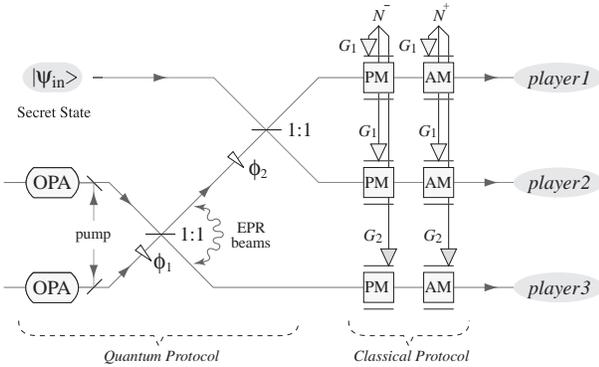}
\caption{ Schematic of the dealer protocol for the 
 $(2,3)$ quantum state sharing scheme. 
  $\psi_{\rm in}$: secret quantum state, OPA:
optical parametric amplifier,  ${\rm
x\!\!:\!\!y}$: beam splitter with reflectivity $x/(x\!+\!y)$ and 
transmitivity $y/(x\!+\!y)$, $\phi_{i}$ optical phase delays,
$N^\pm$: additional Gaussian noise, $G_{i}$ electronic gains.}\label{fig:fig0}
\end{figure}
In this paper we consider quantum states that reside at a the sideband frequency, $\omega$, of an electromagnetic field. These quantum states include the secret and the entangled states use in the dealer protocol, and can be described using the field annihilation operator 
$\hat{a}\!=\!(\hat X^{+}\!+\!i\hat X^{-})/2$.
This operator is expressed in terms of the 
amplitude $\hat X^{+}$ and phase $\hat X^{-}$ quadrature operators, which are non-commuting observables, described by the commutation relation
$[\hat X^{+},\hat X^{-}]=2i$.
Without loss of generality, we can express these quadrature operators in terms of  a steady state component, and fluctuating component as 
$\hat X^{\pm}\!=\!\langle\hat X^{\pm}\rangle\!+\!\delta\hat X^{\pm}$, where 
the variance and the mean of these quadrature operators are expressed as $V^{\pm}\!=\!\langle(\delta\hat X^{\pm})^{2}\rangle$ and 
$\langle\hat X^{\pm}\rangle$, respectively. 
\subsection{Dealer Protocol}
For the $(2,3)$ quantum state sharing scheme, we extend the original dealer protocol proposed by Tyc and Sanders~\cite{Tyc02} ({\it Quantum Protocol} in Fig.~\ref{fig:fig0}). In the original protocol, the secret state is encoded by the dealer by interfering the secret quantum state with one of a pair of EPR entangled beams on a 1:1 beam splitter. This interference hides the secret state in the relatively larger amplitude and phase noise of the entangled beam. The two outputs from this beam splitter, and the second entangled beam, form the three shares to be distributed to the players in the protocol. 
The second EPR entangled beam, although not containing a component of the secret state, does share entanglement with the other two beams. 
This entanglement ensures that the quantum features of the 
secret state can be reconstructed.

The security of the scheme is governed by the strength 
of the entanglement in the dealer protocol. 
In the case of 
finite entanglement, 
some information of the secret state can still be retrieved 
by individual players. 
The security can be further enhanced, however, by using 
additional classical encoding techniques in the dealer protocol. 
This is achieved by encoding correlated Gaussian noise 
onto each of the players shares ({\it Classical Protocol} 
in Fig.~\ref{fig:fig0}). We describe the additional Gaussian 
noise encoded onto the shares by  
$\delta\mathcal{N}\!=\!(\delta{N}^{+}\!+\!i\delta{N}^{-})/2$, 
which has a mean of 
$\langle\delta{N}^{\pm}\rangle\!=\!0$ and variance of $\langle(\delta{N}^{\pm})^{2}\rangle\!=\!V_{N}$. 
The variance of the Gaussian noise 
encoded onto the quadratures of each of the shares 
can be controlled via an electronic gain $G$~(Fig.~\ref{fig:fig0}). 
The resulting shares after the classical encoding can be expressed as ~\cite{Lan03,Lan04}
\begin{eqnarray}
\hat X_{\rm player 1}^{\pm} &=& (\hat X_{{\rm in}}^{\pm}\!+\!\hat X_{\rm EPR1}^{\pm}\!+\!\delta N^{\pm}
)/\sqrt2 \label{dealer1}\\
\hat X_{\rm player 2}^{\pm} &=& (\hat X_{{\rm in}}^{\pm}\!-\!\hat X_{\rm EPR1}^{\pm}\!-\!\delta N^{\pm}
)/\sqrt2 \label{dealer2}\\
\hat X_{\rm player 3}^{\pm} &=& \hat X_{\rm EPR2}^{\pm}\!\pm\!\delta N^{\pm} \label{dealer3}
\end{eqnarray}
where we have assumed that the EPR entangled beams are 
generated by interfering an phase and an 
amplitude squeezed beam on a 1:1 beam splitter 
with a relative optical phase shift of $\pi$. 
The quadratures of the EPR entangled beam are given by 
\begin{eqnarray}
\hat X_{\rm 
EPR1}^{\pm}&=&(\hat X_{\rm sqz1}^{\pm}\!+\!\hat X_{\rm sqz2}^{\pm})/\sqrt{2} \\ 
\hat X_{\rm EPR2}^{\pm}&=&(\hat X_{\rm sqz1}^{\pm}\!-\!\hat X_{\rm sqz2}^{\pm})/\sqrt{2}
\end{eqnarray}
where $\hat X_{\rm sqz1}^{\pm}$ and $\hat X_{\rm sqz2}^{\pm}$ 
correspond to the quadratures of the squeezed beams.  
The variance of the squeezed quadratures of the squeezed 
beams are expressed as 
$V^{-}_{\rm sqz1}\!=\!\langle(\delta\hat X^{-}_{\rm sqz1})^{2}\rangle\!<\!1$
and $V^{+}_{\rm sqz2}\!=\!\langle(\delta\hat X^{+}_{\rm sqz2})^{2}\rangle\!<\!1$.
\subsection{Reconstruction Disentangling Protocols}
\begin{figure}[ht]
\includegraphics[width=8cm]{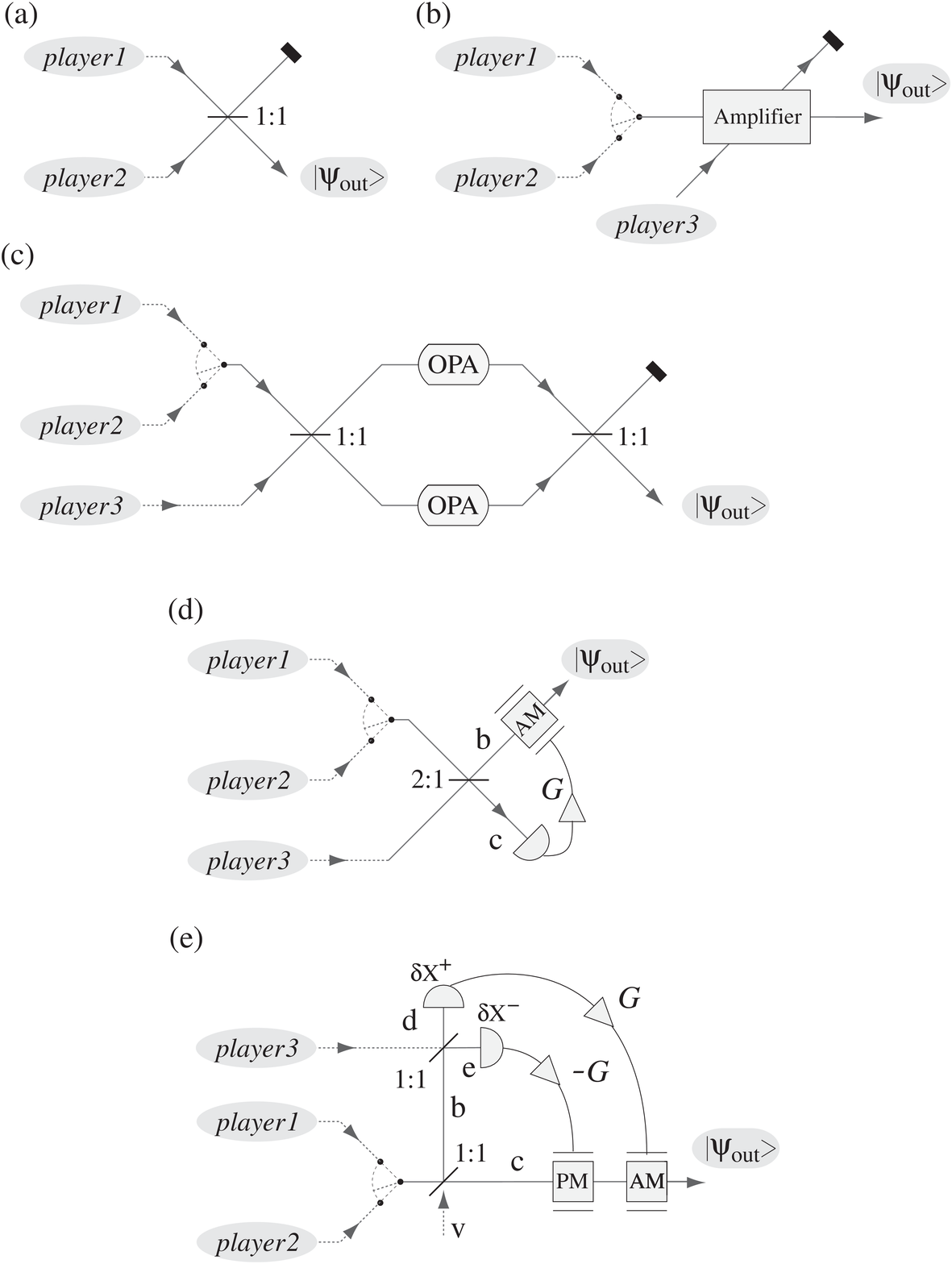}
\caption{ Schematic of the reconstruction protocols for the 
$(2,3)$ quantum state sharing scheme. 
For the players 1 and 2 as the authorized group: (a) Mach-Zehnder reconstruction protocol. 
For the players 1 and 3 (or players 2 and 3) as the authorized group: (b) Phase insensitive amplifier reconstruction protocol. (c) Two optical parametric amplifier reconstruction protocol (d) Feed-forward reconstruction protocol (e) Two feed-forward reconstruction protocol. 
$\psi_{\rm out}$: reconstructed quantum state, 
$G$: electronic gain and  AM: amplitude modulator. Switch symbol: represents the use of either player 1 or player 2 in the corresponding reconstruction protocols. }\label{fig:fig1}
\end{figure}
For $(2,3)$ quantum state sharing, which is the simplest 
non-trivial quantum state sharing scheme, there exists a 
class of protocols that can be used to reconstruct the secret state. 
These reconstruction protocols can be thought of as {\it disentangling protocols} 
as the secret state is embedded within 
two states which are EPR entangled. This is 
analogous to the disentangling protocols in the 
discrete regime \cite{Buz00}. Some of these reconstruction 
protocols are shown in Figure~\ref{fig:fig1}. 
\subsubsection{Mach-Zehnder Protocol}
The specific state reconstruction protocol used in the $(2,3)$ quantum state sharing scheme depends on the constituent players which form the authorized group. 
The authorized group formed when players 1 and 2 collaborate, which we henceforth denote as the \{1,2\}, can reconstruct the secret state using 
the {\it Mach-Zehnder protocol}, as shown in Figure~\ref{fig:fig1}~(a). The  \{1,2\} authorized group completes a Mach-Zehnder interferometer by interfering the shares on a 1:1 beam splitter. The quadratures of the reconstructed secret are given by
\begin{equation}
\hat{X}_{\rm out}^{\pm}\!=\!\frac{\hat{X}_{\rm player 1}^{\pm}\!+\!\hat{X}_{\rm player 2}^{\pm}}{\sqrt{2}}\!=\!\hat{X}_{\rm in}^{\pm}
\end{equation}
Since this reconstruction protocol effectively reverses the dealer encoding protocol, 
the \{1,2\} authorized group can reconstruct the 
secret state to an arbitrary precision, independent of the amount of 
squeezing or additional Gaussian noise employed in the dealer protocol.

For the \{1,3\} and \{2,3\} authorized groups, 
more complicated reconstruction protocols are required. 
This complexity is due to the asymmetry of the entanglement and Gaussian noise in each of the shares in the authorized groups. 

\subsubsection{Phase insensitive amplifier protocol}

The \{1,3\} and \{2,3\} authorized groups can, in theory, reconstruct
the secret state to an arbitrary precision using a phase insensitive 
amplifier. This can be achieved by carefully controlling the inherent 
noise coupled into the output state as a result of the amplification 
process,  as shown in Figure~\ref{fig:fig1}~(b).  The output from a 
phase insensitive amplifier can be expressed by the quadrature 
equations~\cite{Cav82}
\begin{equation}
\hat X^{\pm}_{{\rm out}}=\sqrt{G} \hat  X^{\pm}_{\rm in}\mp\sqrt{G-1}\hat X^{\pm}_{\rm N}
\end{equation}
where $G$ is the amplifier gain. The amplification process has two inputs $\hat  X^{\pm}_{\rm in}$ and $\hat X^{\pm}_{\rm N}$. The second input $\hat X^{\pm}_{\rm N}$ is coupled into the quadratures of the output state as a result of the amplification process. Typically, this second input corresponds to the quadratures of a vacuum state. For a general phase insensitive amplifier, however, this second input can be arbitrary. We utilize this second input in the phase insensitive amplifier reconstruction protocol. In this protocol, share \{1\} (or \{2\} depending on the corresponding authorized group) is amplified using the phase insensitive amplifier. By replacing the amplifier noise coupled into the output state with share \{3\}, the resulting quadratures of the reconstructed secret can be expressed as 
\begin{eqnarray}
\hat X^{\pm}_{{\rm out}}&=&\sqrt{\frac{G}{2}}\hat X^{\pm}_{{\rm in}}+
\Bigg(\frac{\sqrt{G}}{2}\mp\sqrt{\frac{G-1}{2}}\Bigg)\delta\hat X^{\pm}_{{\rm sqz_1}} \nonumber \\
&&+\Bigg(\frac{\sqrt{G}}{2}\pm\sqrt{\frac{G-1}{2}}\Bigg)\delta\hat X^{\pm}_{{\rm sqz_2}} \nonumber \\
&&+\Bigg(\sqrt{\frac{G}{2}}-\sqrt{G-1}\Bigg)\delta N^{\pm}
\end{eqnarray}
By setting the amplifier gain to $G=2$, the quadratures of the reconstructed secret are given by  
\begin{eqnarray}
\hat X^+_{{\rm out}} &=& \hat{X}^+_{{\rm in}}\!+\!\sqrt{2}\delta\hat{X}^+_{\rm sqz_2}\\
\hat X^-_{{\rm out}} &=& \hat{X}^-_{{\rm in}}\!+\!\sqrt{2}\delta\hat{X}^-_{\rm sqz_1}
\end{eqnarray}
where $\delta\hat{X}^+_{\rm sqz_2}$ and $\delta\hat{X}^-_{\rm sqz_1}$ 
are the squeezed quadratures of the squeezed beams used to generate 
the entanglement in the dealer protocol. At an amplifier gain of $G=2$,  
it is seen that in the limit of infinite squeezing, the secret state is 
reconstructed to an arbitrary precision without degradation.
We term this point the {\it unity gain} point 
(analogous to the unity gain point in quantum teleportation \cite{Bow03b,Fur98}).
At the unity gain point the expectation value of the quadrature
amplitudes of the reconstructed state is the same as the secret state. 
Furthermore, the noise contributions do not appear on
the quadratures of the reconstructed state. As a result, 
the security of the scheme can be arbitrarily increased
by either increasing the squeezing or the 
additional noise in the dealer protocol, 
without degrading the quality of the reconstructed state. 
We point out the significance of the 
unity gain point, as all the reconstructions protocols 
presented in this paper can achieve some form of unity gain. 
Although in theory, this reconstruction protocol can be used 
to reconstruct the secret to an arbitrary precision, experimentally 
it is extremely difficult to directly access the second input 
field of the phase insensitive amplifier. 
We now turn our attention to examining more experimentally 
achievable reconstruction protocols. 

\subsubsection{Two optical parametric amplifier protocol}

In their original proposal, Tyc and Sanders 
suggested using a pair of optical parametric amplifiers to perform the 
\{1,3\} and \{2,3\} secret reconstructions~\cite{Tyc02}, as shown in Figure~\ref{fig:fig1}(c).   We term this protocol the 
{\it two optical parametric amplifier protocol}. In this protocol, 
the two shares are interfered on a 1:1 beam splitter. 
The two resulting beams are each noiselessly amplified using phase 
sensitive optical parametric amplifiers, 
with amplifying gains of $\sqrt{G}$ 
and  $1/\sqrt{G}$ respectively. 
After the noiseless amplification, the secret state is reconstructed 
by interfering the two amplified beams on a second 1:1 beam splitter.
 The quadrature of the reconstructed secret can be expressed by 
\begin{eqnarray} \label{eqn:23reconstruction}
\hat X^\pm_{{\rm out}}& =& \frac{1}{2\sqrt{2}}\big(\sqrt{G}-\frac{1}{\sqrt{G}}\big)\hat{X}^\pm_{{\rm
in}}\!+\!\sqrt{2} C^+ \delta N^{\pm} \nonumber \\
&& +C^\pm \delta\hat{X}^\pm_{\rm sqz_1}\!+\!
C^\mp \delta\hat{X}^\pm_{\rm sqz_2}\\
\end{eqnarray}
where $G$ is the amplifying gain of the optical parametric amplifiers, 
and the coefficients $C^{\pm}$ are given by   
\begin{equation}
C^{\pm}=\frac{1}{4}\Big(\frac{1\pm\sqrt{2}}{\sqrt{G}}+(1\mp\sqrt{2})\sqrt{G}\Big)\\
\end{equation}
At an amplifying unity gain of $G=(\sqrt{2}+1)/(\sqrt{2}-1)$, the 
quadratures of the reconstructed state can be expressed as 
\begin{eqnarray}
\hat X^+_{{\rm out}} &=& \hat{X}^+_{{\rm in}}\!+\!\sqrt{2}\delta\hat{X}^+_{\rm sqz_2}\\
\hat X^-_{{\rm out}} &=& \hat{X}^-_{{\rm in}}\!+\!\sqrt{2}\delta\hat{X}^-_{\rm sqz_1}
\end{eqnarray}
In the limit of infinite squeezing in the dealer protocol and at unity gain, 
the secret state is reconstructed to an arbitrary precision. 
This scheme requires significant quantum resources, however, 
with two optical parametric amplifiers in the reconstruction protocol. 
Furthermore, in the reconstruction protocol these optical parametric amplifiers must have precisely controlled amplifying gains as well as high non-linearity. Experimentally, high non-linearity can be achieved by using high peak power pulsed light sources, either in Q-switched or mode-locked setups, or by enhancing the optical intensity within an optical resonator. However, both of these techniques cause significant coupling of vacuum fields into the output state, 
resulting in a significant decrease of quantum efficiency. 
The pulsed systems often suffer distortion of optical wave fronts in the non-linear medium, resulting in poor optical interference and losses, whilst the resonators couple in vacuum fields via intra-resonator losses, the resonator mirrors and the second harmonic pump field. 
For these reasons it is desirable to investigate reconstruction protocols that do not rely on optical parametric amplifiers, but instead utilize linear optics, which are not susceptible to these type of losses and inefficiencies.  

\subsubsection{Single feed-forward reconstruction protocol}

An alternative reconstruction protocol that uses linear optics 
and electro-optic feed-forward to reconstruct the secret 
state~\cite{Lan03,Lan04} is shown in Figure~\ref{fig:fig1}~(d). 
We term this protocol the {\it single feed-forward reconstruction
protocol}. In this protocol the shares are interfered 
on a beam splitter, where 
the two resulting output beams are 
denoted  as $\hat{b}$ and $\hat{c}$. 
The proportion of entanglement and additional noise
between share \{3\} and share \{1\} (or \{2\}) are not equal, 
hence, an appropriate 
beamsplitter ratio must be chosen so that the entanglement and additional noise
contributions are proportional on one of the quadratures of the beam splitter 
output $\hat{b}$ as a result of this interference. 
In the limit of infinite squeezing or additional noise in the 
dealer protocol, the optimum beam splitter ratio is 2:1 
(for convenience we will use this beam splitter ratio 
for the rest of this analysis unless otherwise stated). 
This interference
reconstructs the phase quadrature of the secret state
on the phase quadrature 
of beam splitter output $\hat{b}$. 
As a result of this interference 
the amplitude quadrature $\hat{X}^{+}_{\rm b}$ obtains 
additional noise fluctuations. It is possible to 
cancel the noise on the amplitude quadrature $\hat{X}^{+}_{\rm b}$, 
however, by recognizing that this noise is correlated with the noise 
on the amplitude quadrature $\hat{X}^{+}_{\rm c}$. By detecting 
$\delta\hat{X}^{+}_{\rm c}$ and imparting these fluctuations onto $\hat{X}^{+}_{\rm b}$ 
with a well chosen electronic gain $G$ via an electro-optic 
feed-forward loop, it is possible to reconstruct
the amplitude quadrature of the secret state. 
After the electro-optic feed-forward, 
the quadratures of the reconstructed secret can 
be expressed as 
\begin{eqnarray} \label{eqn:23reconstruction}
\hat X^+_{{\rm out}} &=& \!g^{+}\hat{X}^+_{{\rm
in}}\!+
\!\sqrt{\frac{3}{2}}(\sqrt{3}g^{+}\!-\!1)\delta\hat{X}^+_{\rm
sqz_2}\nonumber\\
& & +\!\frac{1}{\sqrt{2}}(\sqrt{3}\!-\!g^{+})\delta\hat X^-_{\rm
sqz_1}\!+\!(\!\sqrt{3}\!-\!g^{+}\!)\delta N^+\\
\hat X^-_{{\rm out}} &=& \!\frac{1}{\sqrt{3}} (\hat
X^-_{{\rm in}}\!+\!\sqrt{2}\delta\hat{X}^-_{\rm sqz_1})
\end{eqnarray}
where $g^{\pm}$ denotes the optical 
quadrature gains of the reconstructed secret, given by  
$g^{\pm}\!=\!\langle\hat X^{\pm}_{{\rm
out}}\rangle /\langle\hat X^{\pm}_{{\rm in}}\rangle$.
The phase quadrature gain
is set by the reconstruction beam splitter ratio 2:1 to be $g^{-}\!=\!1/\sqrt{3}$, 
whilst the amplitude quadrature gain $g^{+}\!=\!1/\sqrt{3}+G/\sqrt{6}$ 
has an additional contribution due to the
feed-forward process, which is a function of the electronic gain $G$. 
We refer to the optical quadrature gain product 
$g^{+}g^{-}\!=\!(\sqrt{3})(1/\sqrt{3})\!=\!1$ 
as the {\it unity gain} point. At unity gain, the 
quadratures of the 
reconstructed secret are given by
\begin{eqnarray}\label{IdealSSA}
\hat X^+_{{\rm out}} &=& \sqrt{3} (\hat
X^+_{{\rm in}}\!+\!\sqrt{2}\delta\hat{X}^+_{\rm sqz_2})\\
\label{IdealSSB}
\hat X^-_{{\rm out}} &=& \!\frac{1}{\sqrt{3}} (\hat
X^-_{{\rm in}}\!+\!\sqrt{2}\delta\hat{X}^-_{\rm sqz_1})
\end{eqnarray}
In the limit of infinite
squeezing in the dealer protocol, the reconstructed secret is
directly related to the secret state via a local unitary
parametric operation.
This reconstructed state can only be achieved using quantum
resources.  
If required, 
a reverse local unitary parametric operation, can be applied 
to the reconstructed state to transform the reconstructed 
secret state into the original form of the secret state. 
Since this unitary parametric 
operation is a local operation and 
requires no entanglement, this shows the quantum nature 
of the state reconstruction is contained within the feed-forward 
reconstruction protocol, and not by 
the subsequent operations.  

\subsubsection{Double feed-forward reconstruction protocol}

Although the single feed-forward protocol is sufficient for 
demonstrating the quantum nature of quantum state sharing, 
it could be inconvenient in practice if the reconstructed state 
is a unitary transform of the secret state. It is useful to 
investigate alternative feed-forward protocols where the 
reconstructed state is in the same form as the secret state. 
Such a reconstruction protocol is shown in Figure~\ref{fig:fig1}~(e), 
which we term  the {\it double feed-forward reconstruction protocol}. 
In this protocol, share \{1\} (or \{2\}) is interfered with a vacuum 
state on a beam splitter. The reflectivity of the beam splitter 
has to be optimized, and in the limit 
of infinite squeezing or additional 
Gaussian noise in the dealer protocol, the optimum beam 
splitter ratio is 1:1 (For convenience we will use this beam splitter ratio 
for the rest of this analysis unless otherwise stated). 
The resulting output beam $\hat{b}$ is then interfered 
with share \{3\} on a 1:1 beam splitter. The resulting beams 
are denoted by $\hat{d}$ and $\hat{e}$ respectively. The 
noise fluctuations on the amplitude quadrature 
$\hat X^+_{{\rm d}}$ and phase quadrature 
$\hat X^-_{{\rm e}}$ are correlated with the 
amplitude and phase quadrature fluctuations on beam $\hat{c}$
respectively. 
The secret state can be reconstructed by measuring the  
$\hat X^+_{{\rm d}}$ and $\hat X^-_{{\rm e}}$ quadrature 
fluctuations, and displacing the 
corresponding quadratures of beam $\hat{c}$ with a 
properly chosen electronic gain. The resulting quadratures of 
the reconstructed state are given by
\begin{eqnarray} \label{eqn:23reconstruction}
\hat X^{\pm}_{{\rm out}} &=& \!g^{+}\hat{X}^{\pm}_{{\rm
in}}\!+(1\!-\!g^{+})\delta N^+
\!+\!\frac{1}{\sqrt{2}}(1\!-\!g^{+})\hat{A}^{\pm} \nonumber\\
&&+\!\frac{1}{\sqrt{2}}(3g^{+}\!-\!1)\hat{B}^{\pm}
\!+\!\sqrt{2}(g^{+}\!-\!1) \delta\hat{X}^{\pm}_{\rm v} 
\end{eqnarray}
where the coefficients represent the corresponding squeezing operators $\hat{A}^{+}\!=\!\delta\hat{X}^+_{\rm sqz_1}$, $\hat{A}^{-}\!=\!\delta\hat{X}^-_{\rm sqz_2}$, $\hat{B}^{+}\!=\!\delta\hat{X}^+_{\rm sqz_2}$ and 
$\hat{B}^{-}\!=\!\delta\hat{X}^-_{\rm sqz_1}$, the 
optical quadrature gains are defined as $g^{\pm}\!=(1\!-\!G/\sqrt{2})/2$, and where $\delta\hat{X}^\pm_{\rm v}$ is the vacuum noise. At unity gain ($g^{\pm}\!=\!1$),  the quadratures of the reconstructed secret state can be expressed as 
\begin{eqnarray}
\hat X^+_{{\rm out}} &=& \hat{X}^+_{{\rm in}}\!+\!\sqrt{2}\delta\hat{X}^+_{\rm sqz_2}\\
\hat X^-_{{\rm out}} &=& \hat{X}^-_{{\rm in}}\!+\!\sqrt{2}\delta\hat{X}^-_{\rm sqz_1}
\end{eqnarray}
In the case of infinite squeezing in the dealer protocol and at unity gain, the secret state is reconstructed to an arbitrary precision. This protocol has advantages over the previous protocols as it uses linear optics to reconstruct the secret state and the reconstructed state is in the same form as the secret state. 
\section{Characterization of Quantum State Reconstruction}
We characterize state reconstruction in quantum state sharing by measuring 
the fidelity between the secret and reconstructed states $(\mathcal{F})$, 
which is used in the characterization quantum 
teleportation experiments~\cite{Fur98, Bow03b}. We also characterize the 
state reconstruction by measuring the 
signal transfer from the secret to the reconstructed state $(\mathcal{T})$ and the additional noise on the reconstructed state $(\mathcal{V})$, which is used to characterize quantum teleportation~\cite{Bow03b} and quantum non-demolition experiments~\cite{Poi94}.
\subsection{Fidelity}
Fidelity measures the overlap between the secret and reconstructed states, and can be expressed in terms of the input and the output state as  $\mathcal{F}=\langle\psi_{\rm in} \big{|}\hat{\rho}_{\rm out}\big{|}\psi_{\rm in}\rangle$~\cite{Shu95}. A fidelity of $\mathcal{F}\!=\!1$, implies
perfect overlap between the secret and reconstructed states and corresponds to state reconstruction with arbitrary precision, whilst a fidelity of $\mathcal{F}\!=\!0$ implies no overlap between the corresponding states.

In quantum state sharing the secret state can be any state in general. 
In our experiment we use coherent states with unknown amplitude and phase
coherent amplitudes. As a consequence we limit our analysis here to coherent states. 
The fidelity between a secret state and the reconstructed state for a general quantum state sharing scheme, assuming that all states have Gaussian statistics, can be expressed as
\begin{equation}\label{Fidelity}
\mathcal{F}=\frac{2{\rm e}^{-(k^{+}\!+\!k^{-})/4}}{\sqrt{(1\!+\!V_{{\rm
out}}^{+})(1\!+\!V_{{\rm out}}^{-})}}
\end{equation}
where $k^{\pm}\!=\!\langle X_{{\rm
in}}^{\pm}\rangle^{2}(1\!-\!g^{\pm})^{2}/ (1\!+\!V_{{\rm
out}}^{\pm})$ and $V_{{\rm out}}^{\pm}$ are the 
quadrature variances of the reconstructed state, and where $g^{\pm}$ are the optical quadrature gains.   
Since fidelity is a measure of the overlap between the input and the output state, the most significant fidelity measure is at unity gain $g^{\pm}=1$. This is seen as the fidelity, averaged over an ensemble of unknown states, falls exponentially as we move away from unity gain.   

We now determine the maximum fidelity achievable by the authorized group in the case 
when the squeezed states are replaced with coherent states in the dealer protocol,  
which we term the classical fidelity limit. 
The quadrature equations for a
general reconstructed state
can be expressed as 
\begin{equation} \label{aa}
\hat{X}^{\pm}_{\rm out}\!=\!g^{\pm}\hat{X}^{\pm}_{\rm in}\!+\!\hat{X}^{\pm}_{\rm N}
\end{equation}
where $\hat{X}^{\pm}_{\rm N}$ are the reconstruction noise terms on the quadratures of the reconstructed state. To measure the fidelity of this reconstructed state at unity gain, we assume that a phase insensitive amplification can be applied to the reconstructed state to achieve unity gain. Assuming that the optical quadrature gains on both quadratures are equal $g^{\pm}\!=\!g$, and for a phase insensitive amplification with an amplifying gain of $1/g$, the resulting quadrature equations are given by
\begin{equation} \label{ab}
\hat{X}^{\pm}_{\rm out(amp)}\!=\!\hat{X}^{\pm}_{\rm in}\!+\!\frac{1}{g}\hat{X}^{\pm}_{\rm N}\!+\!\hat{X}^{\pm}_{\rm M}
\end{equation}
where $\hat{X}^{\pm}_{\rm M}$ is the noise coupled into the output state as a result of the amplification process. By using the commutation relation 
$[X^{+}_i,X^{-}_j]\!=\!2i\delta_{ij}$ and the Heisenberg uncertainty product inequality 
$V^{+}_i V^{-}_j \!\geq\! |\langle[X^{+}_i,X^{-}_j]\rangle|^2/4$, we can obtain Heisenberg uncertainty products for the noise terms in Equations (\ref{aa}) and (\ref{ab}) expressed as
\begin{equation}  \label{ba}
V^{+}_{\rm N}V^{-}_{\rm N}\!\geq\!{|(1\!-\!g^{2})|}^2
\end{equation}
\begin{equation} \label{bb}
V^{+}_{\rm M}V^{-}_{\rm M}\!\geq\!{|(1\!-\!g^{2})/(g^{2})|}^2
\end{equation}
By substituting Equations~(\ref{ab}),~(\ref{ba}) and~(\ref{bb}) into fidelity Equation~(\ref{Fidelity}), 
the maximum classical fidelity for a general reconstructed state is given by
\begin{equation} \label{bc}
\mathcal{F}^{\rm clas}\leq\frac{1}{1+|(1\!-\!g^{2})/g^{2}|}
\end{equation}
Using this inequality, we can determine the maximum classical fidelity achievable by the authorized groups for the $(2,3)$ quantum state sharing scheme. From the  
individual player shares, Equations~(\ref{dealer1}), (\ref{dealer2}) and (\ref{dealer3}), the 
quadrature gains for \{1,2\} access group are $g^{\pm}\!=\!1$, whilst the quadrature gains for \{1,3\} and \{2,3\} access group are $g^{\pm}\!=\!1/\sqrt{2}$. By substituting these gains into Equation~(\ref{bc}), the maximum classical fidelity for the authorized groups are given by
\begin{eqnarray}
\mathcal{F}^{\rm clas}_{\{1,2\}} &\leq &1 \nonumber \\ 
\mathcal{F}^{\rm clas}_{\{1,3\}}&=&\mathcal{F}^{\rm clas}_{\{2,3\}} \leq 1/2 
\end{eqnarray}
The average classical fidelity limit for the quantum state sharing 
scheme can be determined by averaging the maximum classical 
fidelity achievable by all the authorized groups.
For the $(2,3)$ quantum state sharing scheme, the average 
classical fidelity is $\mathcal{F}^{\rm clas}_{\rm avg}\!\leq\!(\mathcal{F}_{\{1,2\}}\!+\!\mathcal{F}_{\{1,3\}}\!+\!\mathcal{F}_{\{2,3\}})/3\!=\!2/3$. This limit can only be exceeded 
using quantum resources. The average classical fidelity 
achievable for a general $(k,n)$ quantum state sharing 
scheme can also be calculated. Assuming that the secret 
is a coherent state it is straightforward to show that the 
average classical fidelity is given by 
$\mathcal{F}^{\rm clas}_{\rm avg}\!\leq\!k/n$.

Similarly for the individual players, the maximum achievable
classical fidelity limits are given by
\begin{eqnarray}
\mathcal{F}^{\rm clas}_{\{1\}}&=&\mathcal{F}^{\rm clas}_{\{2\}} \leq 1/2  \nonumber \\   
\mathcal{F}^{\rm clas}_{\{3\}} &= &0
\end{eqnarray}
For large squeezing or additional noise in the 
dealer protocol, the fidelity for the individual players
approaches zero, corresponding to no overlap between the 
secret state and the individual shares.

\subsection{Signal Transfer and Additional Noise}

In quantum state sharing, the state reconstruction can also be characterized in terms of
the signal transfer to ($\mathcal{T}$) and additional noise on 
($\mathcal{V}$) the reconstructed state. These measures 
provides complementary information about the state reconstruction compared
with the fidelity measure. 
Perfect state reconstruction corresponds to $\mathcal{T}\!=\!2$ and 
$\mathcal{V}\!=\!0$, whilst $\mathcal{T}\!=\!0$ and 
$\mathcal{V}\!=\!\infty$ implies that no 
information has been obtained about the secret state. 
The spacial difference between the 
$\mathcal{T}$ and $\mathcal{V}$ points, 
for the access and adversary groups, illustrates the 
information difference about the secret state obtained by both groups.
Unlike fidelity, which requires the reconstructed state
to be in the same form as the secret state, both $\mathcal{T}$
and $\mathcal{V}$ are 
state independent measures and are 
invariant to unitary transformations of the
reconstructed state. 

The signal transfer function is given by 
the sum of the quadrature signal transfer coefficients $T^{\pm}$ as 
\begin{equation}
\mathcal{T}\!=\!T^{+}+T^{-}\!=\!\frac{{\rm R}^{+}_{\rm out}}{{\rm R}^{+}_{\rm in}}\!+\!
\frac{{\rm R}^{-}_{\rm out}}{{\rm R}^{-}_{\rm in}}
\end{equation} 
where ${\rm R}^{\pm}$ are the quadrature signal-to-noise ratios. 
In the case of zero squeezing in the dealer protocol, using Equation~(\ref{ba}), 
the signal transfer for a general reconstructed state, Equation~(\ref{aa}), 
is limited by the inequality
\begin{equation} \label{ac}
\mathcal{T}^{\rm clas}\!\leq\!\frac{1} {1+|1/(g^{+})^2-1|}+
\frac{1} {1+|1/(g^{-})^2-1|}
\end{equation}
The additional noise on the reconstructed state ($\mathcal{V}$) is given by product of the quadrature conditional variances, which can be expressed as 
\begin{equation}
\mathcal{V}\!=\!V^{+}_{\rm in|out}V^{-}_{\rm in|out} 
\end{equation}
where the quadrature conditional variances each describe the amount of additional noise on each quadrature of the secret state and can be expressed in the standard form 
$V^{\pm}_{\rm in|out}=\min_{h^{\pm}_{\rm in}}
\langle(\delta\hat{X}^{\pm}_{\rm out}-h^{\pm}_{\rm
in}\delta\hat{X}^{\pm}_{\rm in})^{2}\rangle$, where the gains $h^{\pm}_{\rm in}$ are
optimized, giving minimum conditional variances of 
\begin{equation} \label{CV}
V^{\pm}_{\rm in|out}\!=\!V^{\pm}_{\rm in}\!-\!
\frac{|\langle\delta\hat{X}^{\pm}_{\rm in}\delta\hat{X}^{\pm}_{\rm out}\rangle |^{2}}{V^{\pm}_{\rm out}}\end{equation}
For a general reconstructed state described by Equation~(\ref{aa}), and assuming that the secret is a coherent state, the quadrature conditional variances can be written in an alternative form as $V^{\pm}_{\rm in|out}\!=\!(V^{\pm}_{\rm out}\!-\!(g^{\pm})^{2})$. The minimum additional noise on the reconstructed state is limited by the inequality
\begin{equation}  \label{ad}
\mathcal{V}\!\geq\!|1-g^{+}g^{-}|^{2}
\end{equation}

For our (2,3) quantum state sharing protocol, we determine the classical limits
for $\mathcal{T}$ and $\mathcal{V}$ for the authorized groups. The \{1,2\} authorized group can obtain a maximum signal transfer, and a minimum additional noise of 
\begin{eqnarray}
\mathcal{T}^{\rm clas}_{\{1,2\}} &\leq &2 \nonumber \\ 
\mathcal{V}^{\rm clas}_{\{1,2\}}& \geq &0 
\end{eqnarray}
which corresponds to state reconstruction to an arbitrary precision. 
For the \{1,3\} and \{2,3\} authorized groups, the maximum achievable signal transfer, and the minimum achievable additional noise is given by
\begin{eqnarray} \label{da}
\mathcal{T}^{\rm clas}_{\{1,3\}} &=&\mathcal{T}^{\rm clas}_{\{2,3\}} \leq 1 \nonumber \\ 
\mathcal{V}^{\rm clas}_{\{1,3\}}&=&\mathcal{V}^{\rm clas}_{\{2,3\}} \geq  1/4 
\end{eqnarray}
For no squeezing in the dealer protocol, the \{1\} and \{2\} adversary groups can reach the equality given in Equation~(\ref{da}). As either the squeezing or additional Gaussian noise is increased in the dealer protocol, however, 
the amount of information the adversary group obtains approaches zero. 
In the limit of infinite squeezing, or large amounts of additional noise, the adversary groups obtains no information about the secret state, corresponding to $\mathcal{T}\!=\!0$ and $\mathcal{V}\!=\!\infty$. 

Figures~\ref{fig:fig2a}~and~\ref{fig:fig2b} show the accessible  $\mathcal{T}$ and $\mathcal{V}$ regions for the \{1,3\} and \{2,3\} authorized groups using the single feed-forward reconstruction protocol, with and without squeezing in the dealer protocol. The accessible points for the corresponding  \{2\} and \{1\} adversary groups are also shown. To map out these accessible regions, the authorized group vary both the electronic feed-forward gain, and the beam splitter reflectivity in the reconstruction protocol.  In Figure~\ref{fig:fig2a}, for no squeezing in the dealer protocol, the authorized group can achieve the classical limits set in Equation~(\ref{da}). Figure~\ref{fig:fig2b} shows that in the limit of ideal squeezing, the authorized group can achieve state reconstruction to an arbitrary precision. 
\begin{widetext}
\begin{center}
\begin{figure}[h]
\includegraphics[width=15cm]{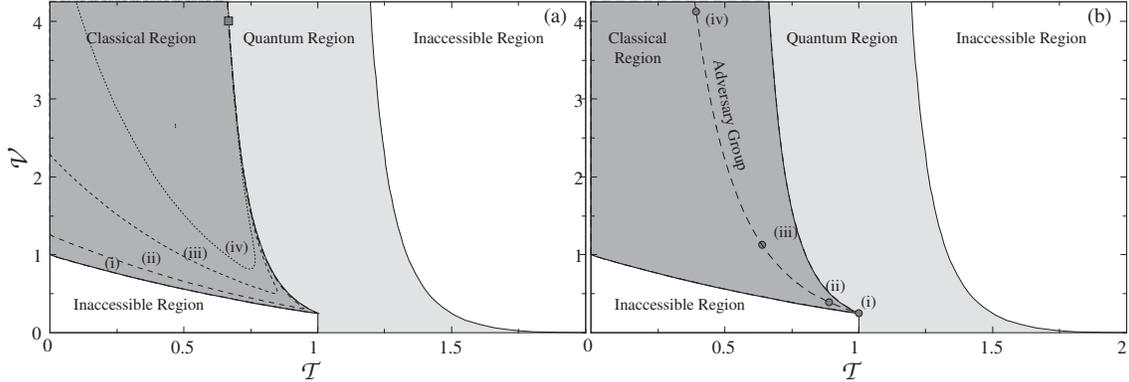}
\caption{Signal transfer $(\mathcal{T})$ and 
additional noise $(\mathcal{V})$ for 
the {\it single feed-forward reconstruction protocol}, 
with no squeezing in the dealer protocol.
(a) Accessible regions for the \{2,3\} (and \{1,3\}) authorized groups, 
and (b) accessible points for the \{1\} (and \{2\}) adversary group, for 
increasing additional Gaussian noise in the dealer protocol of (i) $V_{\rm N}\!=\!0$ (ii) $V_{\rm N}\!=\!0.25$ (iii) $V_{\rm N}\!=\!1.13$ and (iv) $V_{\rm N}\!=\!3.06$, normalized to the quantum noise limit.  
Grey square: unity gain point for the authorized group reconstruction protocol.
Grey circles: corresponding adversary group points. 
}\label{fig:fig2a}
\end{figure}
\begin{figure}[h]
\includegraphics[width=15cm]{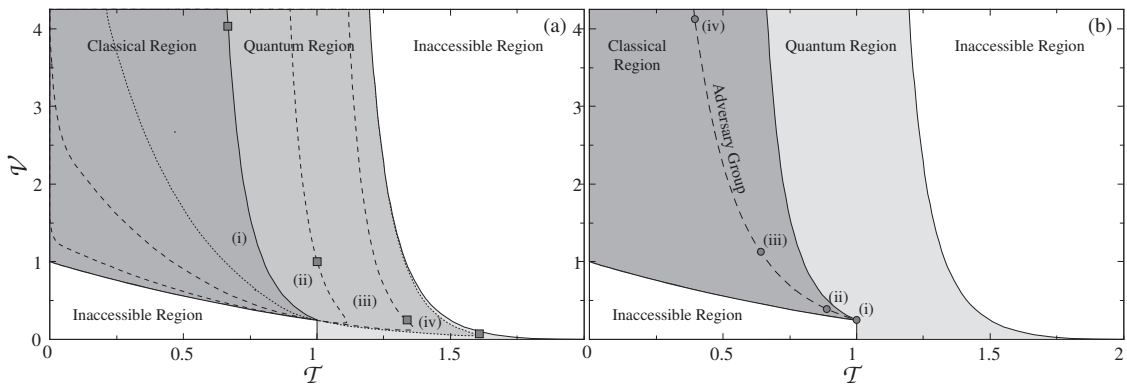}
\caption{ 
Signal transfer $(\mathcal{T})$ and 
additional noise $(\mathcal{V})$ for
the {\it single feed-forward reconstruction protocol}, with increasing squeezing in 
the dealer protocol. 
(a) Accessible regions for the \{2,3\} (and \{1,3\}) authorized groups, 
and (b) accessible points for the \{1\} (and \{2\}) adversary group, for 
increasing squeezing in the dealer protocol of (i) $0$~dB (ii) $-3$~dB (iii) $-6$~dB and (iv) $-9$~dB below the quantum noise limit, and with no additional Gaussian noise. 
}\label{fig:fig2b}
\end{figure}
\end{center}
\end{widetext}
\section{Experimental Setup} 
\subsection{Dealer Protocol}
We use a Nd:YAG laser producing 1.2~W of laser light at 1064~nm. 
Approximately 0.8~W of this laser light is coupled into  a hemilithic ${\rm MgO\!:\!LiNbO}_{\rm 3}$ second harmonic generator, producing 
approximately 0.4W of frequency doubled light at 532nm.  

The remaining light from the laser is coupled into a high finesse mode cleaning cavity. This cavity 
serves as stable frequency reference, to which the laser is locked. The output beam is  
quantum noise limited above the sideband frequency of 2MHz.
The mode cleaning cavity also ``spatially cleans" the output mode,
by only being resonant for the ${\rm TEM}_{00}$ transverse 
electromagnetic field mode. 
This output beam is used to seed 
two hemilithic ${\rm MgO:LiNbO}_{\rm 3}$ optical parametric 
amplifiers, which are pumped  with the frequency doubled light. 
The optical phase of the pump beam is controlled to produce 
amplitude squeezed beams from the optical parametric amplifiers. 
The amount of amplitude quadrature squeezing corresponds to 
$-4.5\!\pm\!0.2$~dB below the quantum noise limit.
To produce EPR entangled beams, the two amplitude squeezed 
beams are interfered on a 1:1 beam splitter with a controlled 
relative optical phase shift of $\pi/2$. The resulting beams exhibit 
continuous variable entanglement between the amplitude and phase 
quadratures of the two beams. This entanglement is
characterized using two standard measures. The first measure, 
proposed by Duan {et al.}~\cite{Dua00}, characterizes the 
inseparability of the two entangled wave functions and is 
referred to as the {\it inseparability criterion}. Our system 
satisfies the inseparability criterion, which can be express as
\begin{equation}
\sqrt{V^{+}_{\rm EPR1+EPR2} V^{-}_{\rm EPR1-EPR2}}\!=\!0.44\!\pm \!0.01\!<\!1
\end{equation}
where $V_{\rm EPR1\pm EPR2}$ is the minimum of the normalized variance of the sum or difference of the operators $\hat{X}_{\rm EPR1}$ and $\hat{X}_{\rm EPR2}$.
 A second measure proposed by Reid and Drummond~\cite{Rei88},
referred to as the {\it EPR criterion}, is based on the observation of non-classical correlations which can be used to demonstrate the EPR paradox. 
Our system satisfied the EPR criterion which can be express as
\begin{equation}
V^{+}_{\rm EPR1|EPR2} V^{-}_{\rm EPR1|EPR2}\!=\!0.58\!\pm\! 0.02\!<\!1
\end{equation}
where $V^{\pm}_{\rm EPR1|EPR2}$ are the standard conditional 
variances given in Equation~(\ref{CV}). A more detailed analysis and 
discussion of the experimental generation and characterization of 
continuous variable EPR entanglement is given in ~\cite{Bow03}. 

In our experiment, the secret quantum state is a displaced coherent 
state at the sideband frequency of 6.12~MHz of the coherent laser 
field.
The secret state is encoded and distributed to the three players by 
interfering it with one of the EPR entangled beams on a 1:1 beam 
splitter with a mode-matching efficiency of 
$\eta_{\rm EPR1,in}\!=\!0.97$, as shown in Figure~\ref{fig:fig0}. 
To increase the security of the scheme, the dealer introduces
additional Gaussian noise onto the three player shares using 
electro-optic modulation techniques. 
An alternative method for introducing this noise is to modulate 
the optical parametric amplifier resonator cavities with Gaussian 
noise at the secret state sideband frequency. 
In our experiment, the Gaussian noise appears 
naturally as a result of de-coherence in the optical parametric 
amplifiers, resulting in mixed output states from the 
optical parametric amplifiers. These mixed states can be 
described as squeezed states with additional noise on the 
anti-squeezed quadratures. 
The additional noise on the EPR beams corresponds
exactly to Equations (\ref{dealer1}), (\ref{dealer2}) and (\ref{dealer3}). 
Experimentally, the variance of this noise can be controlled 
to an extent by adjusting the power of the 532nm light used 
to pump the optical parametric amplifiers. Typically, the additional 
Gaussian noise has a noise variance of $3.5$~dB above the 
quantum noise limit, for a corresponding amplitude quadrature 
squeezing of $4.5\!\pm\!0.2$~dB below the quantum noise limit.
\subsection{Reconstruction Protocols}
For the \{1,2\} authorized group state reconstruction, we use 
the Mach-Zehnder reconstruction protocol, as shown in 
Figure~\ref{fig:fig1}~(a). Both players shares are interfered 
on a 1:1 beam splitter with a mode-matching efficiency of 
$\eta_{\rm share1,share2}\!=\!0.99$. The output state
 from the beam splitter is the reconstructed secret state. 

For the $(2,3)$ quantum state sharing scheme, the \{1,3\} 
and \{2,3\} authorized groups are equivalent, so that an 
experimental demonstration requires the successful demonstration 
of either of these protocols. For the \{2,3\} authorized group,
 the secret state is reconstructed using the single 
 feed-forward reconstruction protocol, as shown in 
 Figure~\ref{fig:fig1}~(d). For this reconstruction protocol, 
 the players shares are interfered on a 2:1 beam splitter 
with a mode-matching 
 efficiency of $\eta_{\rm player2,player3}\!=\!0.97$.
To improve the efficiency of the feed-forward loop, the optical 
power on the feed-forward detector is increased so that 
the quantum noise limit is sufficiently higher than the detector 
dark noise. Typically for our experiment, the dark noise on the feed-forward 
detector is $13$~dB below the quantum noise limit. 

In the original proposal, the amplitude quadrature of beam 
$\hat{b}$ is directly modulated using electro-optic feed-forward 
techniques. This method, however, is prohibitive as amplitude 
modulators have quantum efficiencies of 50\%. 
An alternative method that has a much higher quantum efficiency 
is to displace the amplitude quadrature of a separate strong local oscillator field $\hat{X}^{+}_{\rm LO}$. 
This local oscillator field is then interfered with field $\hat{b}$ 
on a highly reflective beam splitter. The efficiency of this technique 
is equal to the beam splitter reflectivity of the highly reflective 
beam splitter. In our 
experiment we use a beam splitter ratio of 50:1 with a 
mode-matching efficiency of  $\eta_{\rm share3, LO}\!=\!0.96$. 
%

%
\subsection{Measuring the Secret and Reconstructed Quantum States}
Both the secret and reconstructed quantum states for the 
access and adversary groups are measured using a single 
balanced homodyne detector, via a configuration of 
removable mirrors. Assuming Gaussian states, the secret and 
reconstructed states are completely characterized by 
measuring the amplitude and phase quadrature noise spectra, 
together with the calibration noise spectra.
After detection, the total homodyne efficiency $\eta_{\rm hom}\!=\!0.89\!\pm\!0.01$ is factored into each measurement. This inference ensures accurate results (this can be seen in the limit of poor homodyne efficiency, where all states measured correspond to vacuum states, resulting in state reconstruction to an arbitrary precision by both the access and adversary groups, which would be obviously incorrect).
Due to a control drift in our experimental setup, the quadrature noise spectra are normalized with respect to the noise of the secret state, which are approximately quantum noise limited at 6.12~MHz. From these noise spectra, $\langle\hat{X}^{\pm} \rangle$ and $\langle(\hat{X}^{\pm})^2\rangle$ of the secret and reconstructed states are calculated respectively. 
\section{Experimental Results}
\begin{figure}[ht]
\includegraphics[width=8cm]{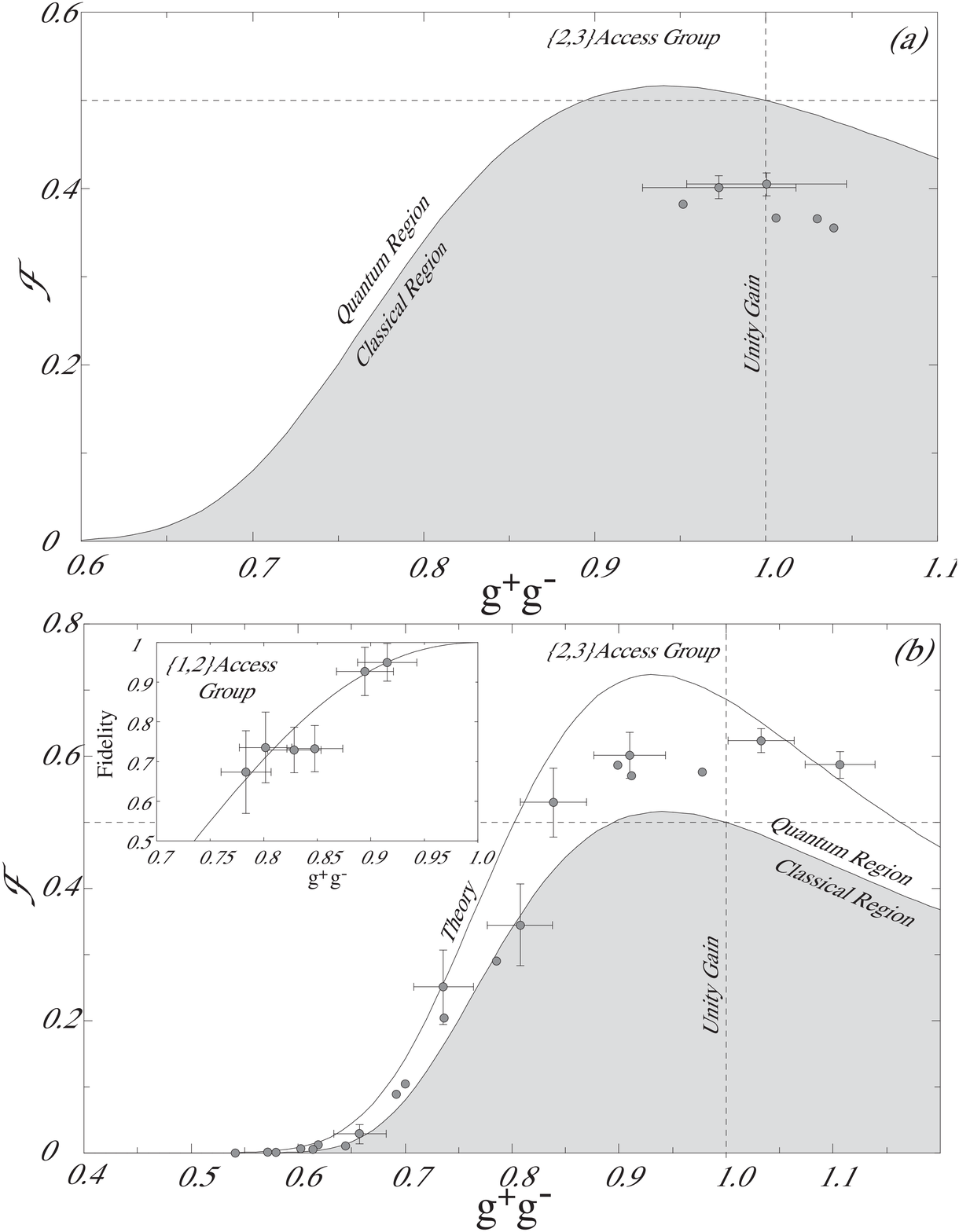}
\caption{ Experimental fidelity for the authorized groups. 
(a) Classical fidelity for \{2,3\} authorized group as a function of the optical gain product $g^{+}g^{-}$.
(b) Fidelity for \{2,3\} authorized group with $4.5$~dB of squeezing in the dealer protocol. Solid line: theoretical curve with $3.5$~dB of additional Gaussian noise, $13$~dB of electronic noise below the quantum noise limit and a feed-forward detector efficiency of $\eta_{\rm ff}\!=\!0.93$. Grey area: classical region for the authorized group. (inset) Fidelity for the \{1,2\} authorized group as a function of the optical gain product. Solid line: theoretical curve. 
 }\label{fig:fig3}
\end{figure}
\subsection{Experimental Fidelity Results}
Figure~\ref{fig:fig3}~(b)~(inset) shows the measured fidelity for the \{1,2\} authorized group as a function of the optical gain product $g^{+}g^{-}$,
with $-4.5\!\pm\!0.2$~dB of squeezing and $+3.5\!\pm\!0.1$~dB of additional Gaussian noise in the dealer protocol. 
The authorized group obtains a best fidelity of $\mathcal{F}_{\{1,2\}}\!=\!0.95\!\pm\!0.05$ with $g^{+}g^{-}\!=\!0.92\!\pm\!0.03$. The theoretical curve for the fidelity as a function of optical gain product is also shown.
The fidelity and the optical gain product for the \{1,2\} authorized group are close to unity, being slightly degraded as a result of experimental losses and imperfections.

For the  \{2,3\} and \{1,3\} authorized groups using the single feed forward reconstruction protocol, a meaningful fidelity measure cannot be obtained directly. This is because the reconstructed secret is a unitary transform of the secret state, and the overlap between the secret and reconstructed state overlap is poor, even in the ideal case of infinite squeezing. However, a meaningful fidelity measure can be obtained
after the unitary parametric operation 
$\delta\hat{X}^{\pm}_{{\rm para}}\!=\!  (\sqrt{3})^{\mp
1}\delta\hat{X}^{\pm}_{{\rm out}}$
is applied to reconstructed state {\it a posteriori}. This unitary parametric operation can either be applied electronically to the measured values of the amplitude and phase quadratures of the reconstructed state, or optically by amplifying the reconstructed state using an optical parametric amplifier. 

Figure~\ref{fig:fig3}~(a) shows the measured fidelity for the \{2,3\} authorized group as a function of the optical gain product for zero squeezing in the dealer protocol. 
The \{2,3\} authorized group achieves a maximum fidelity of $\mathcal{F}^{\rm clas}_{\{2,3\}}\!=\!0.41\!\pm\!0.01$ at a gain of $g^{+}g^{-}\!=\!1.00\!\pm\!0.05$. 
Figure~\ref{fig:fig3}~(b) shows the measured fidelity for the \{2,3\} authorized group as a function of the optical gain product, For $-4.5\!\pm\!0.2$~dB of squeezing and $+3.5\!\pm\!0.1$~dB of additional Gaussian noise in the dealer protocol.  Near unity gain of $g^{+}g^{-}\!=\!1.03\!\pm\!0.03$, the \{2,3\} authorized group measures a best state reconstruction of $\mathcal{F}_{\{2,3\}}\!=\!0.62\!\pm\!0.02$. This fidelity exceeds the classical fidelity limit $\mathcal{F}^{\rm clas}_{\{2,3\}}\!\leq\!1/2$, which is only achievable using quantum resources in the dealer protocol. 
In our scheme, the fidelity averaged over all authorized groups is $\mathcal{F}_{\rm avg}\!=\!(\mathcal{F}_{\{1,2\}}\!+\!2\mathcal{F}_{\{2,3\}})/3\!=\!0.73\!\pm\!0.02$, which exceeds the classical limit of $\mathcal{F}^{\rm clas}_{\rm avg}\!=\!2/3$. This classical limit can only be exceeded using quantum resources and so demonstrates the quantum nature of the $(2,3)$ threshold quantum state sharing scheme. 

For the corresponding adversary group \{1\}, the fidelity is calculated both with, and without,  an ideal linear amplification applied to the reconstructed state to achieve unity gain. This linear amplification operation is applied to the measured quadratures of the adversary state electronically after the measurement, and is described by Equation~(\ref{ab}), where we assume a linear amplification gain of $\sqrt{2}$. 
In the case of no squeezing in the dealer protocol,  the best fidelity achieved by the adversary group after amplification is $\mathcal{F}^{\rm clas}_{\{1\} \rm amp}\!=\!0.25\!\pm\!0.01$ with a gain of $2g^{+}g^{-}\!=\!0.92\!\pm\!0.04$, where the subscript (amp) denotes the fidelity after the {\it a posteriori} linear amplification. In this case, however, the adversary group obtains a higher fidelity by not applying linear amplification operation, with a fidelity $\mathcal{F}^{\rm clas}_{\{1\}}\!=\!0.35\!\pm\!0.06$ at a gain of $g^{+}g^{-}\!=\!0.46\!\pm\!0.02$ achieved directly without amplification. 

For $-4.5\!\pm\!0.2$~dB of squeezing and $+3.5\!\pm\!0.1$~dB of additional Gaussian noise in the dealer protocol,  the best fidelity achieved by the adversary group is $\mathcal{F}_{\{1\} \rm amp}\!=\!0.16\!\pm\!0.01$ at a gain of $2g^{+}g^{-}\!=\!1.00\!\pm\!0.04$.  Without the linear amplification the authorized group achieves a fidelity of $\mathcal{F}_{\{1\}}\!=\!0.04\!\pm\!0.02$ at a gain of $g^{+}g^{-}\!=\!0.50\!\pm\!0.02$

\begin{figure}[ht]
\includegraphics[width=8cm]{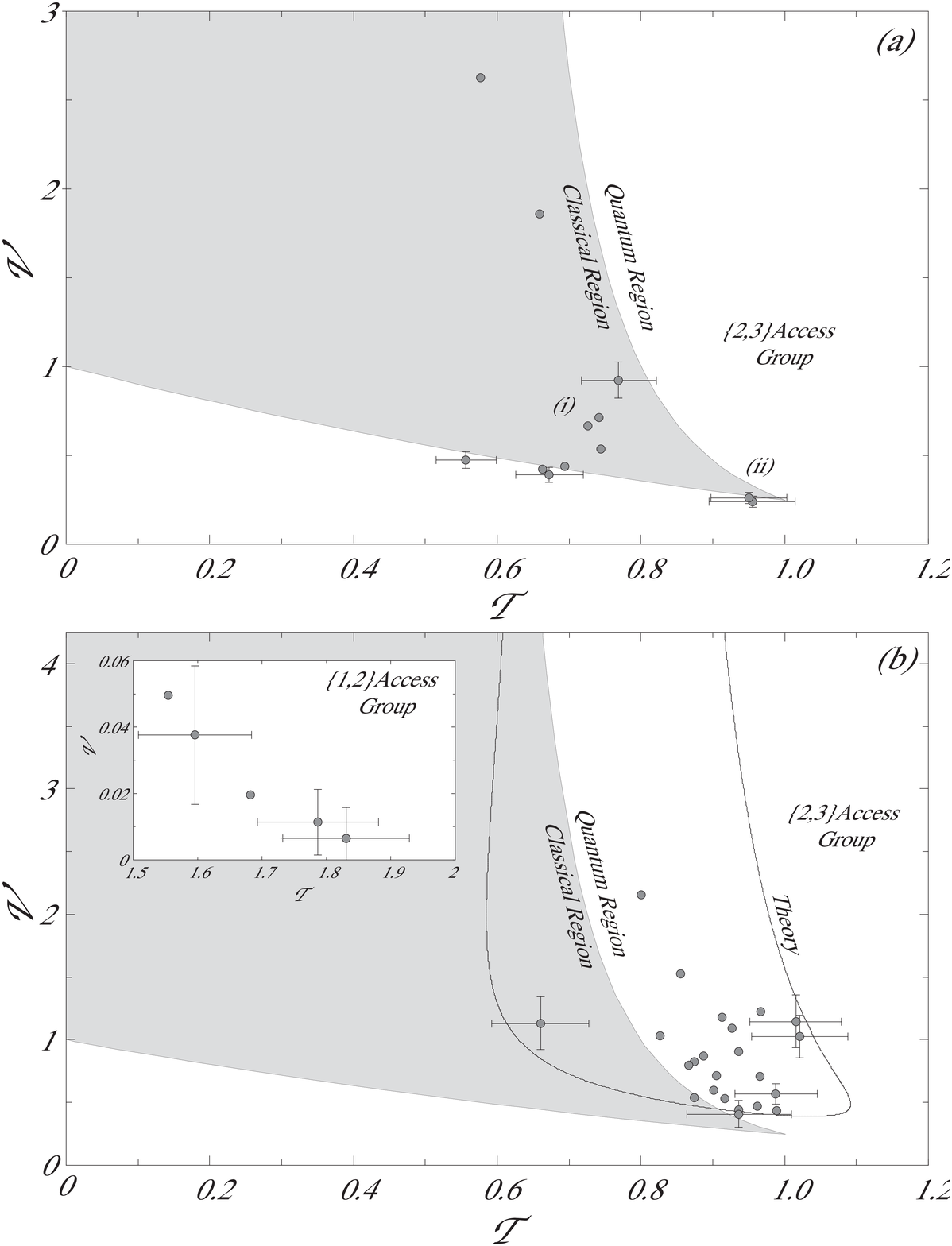}
\caption{Experimental signal transfer $(\mathcal{T})$ and 
additional noise $(\mathcal{V})$ for the authorized groups. 
(a) Classical $\mathcal{T}$ and  $\mathcal{V}$ for the \{2,3\} authorized group, for varying electronic feed-forward gain. 
(b) $\mathcal{T}$ and  $\mathcal{V}$ for the \{2,3\} authorized group with $-4.5$~dB of squeezing in the dealer protocol. Solid line: theoretical curve for authorized group. Grey area: classical region for the authorized group. (inset) Experimental $\mathcal{T}$ and $\mathcal{V}$ for the \{1,2\} authorized group. 
}\label{fig:fig4}
\end{figure}
\subsection{Experimental $\mathcal{T}$ and $\mathcal{V}$ Results}
Figure~\ref{fig:fig4}~(a)~(inset) shows the measured signal transfer $\mathcal{T}$ and additional noise $\mathcal{V}$ for the \{1,2\} authorized group with $-4.5\!\pm\!0.2$~dB of squeezing and $+3.5\!\pm\!0.1$~dB of additional Gaussian noise in the dealer protocol. 
The \{1,2\} authorized group achieves a best state reconstruction of $\mathcal{T}_{\{1,2\}}\!=\!1.83\!\pm\!0.10$ and 
$\mathcal{V}_{\{1,2\}}\!=\!0.01\!\pm\!0.01$, which are both close to 
$\mathcal{T}\!=\!2$ and $\mathcal{V}\!=\!0$, corresponding to 
ideal state reconstruction. 

Figure~\ref{fig:fig4}~(b) shows the $\mathcal{T}$ and $\mathcal{V}$ points for 
the \{2,3\} authorized group for no squeezing or additional Gaussian noise in the dealer protocol. The points are taken from two experimental runs. 
The first experimental points (labeled (i)) are for a reconstruction beam splitter ratio of 2:1 and for varying electronic feed-forward gain. The second experimental points (labeled (ii)) are for an optimized beam-splitter reflectivity of 100\% and for zero electronic feed-forward gain. In this case the \{2,3\} authorized group obtains a best signal transfer of $\mathcal{T}^{\rm clas}_{\{2,3\}}\!=\!0.96\!\pm\!0.06$ and lowest additional noise of $\mathcal{V}^{\rm clas}_{\{2,3\}}\!=\!0.24\!\pm\!0.03$.  These points are close to the classical limits described by Equations~(\ref{ac})~and~(\ref{ad}).

Figure~\ref{fig:fig4}~(b) shows the measured $\mathcal{T}$ and $\mathcal{V}$ points for the \{2,3\} authorized group for $-4.5\!\pm\!0.2$~dB of squeezing and $+3.5\!\pm\!0.1$~dB of additional Gaussian noise in the dealer protocol. The points are taken for 
a varying electronic feed-forward gain, and a beam splitter ratio of 2:1. 
The classically accessible region, which can only be exceeded using quantum resources in the dealer protocol, is also shown. The quantum nature of our protocol is demonstrated by the experimental points which exceed this classical region.  For the \{2,3\} authorized group, we measure a best signal transfer of 
$\mathcal{T}_{\{2,3\}}\!=\!1.01\!\pm\!0.06$ and lowest additional noise of $\mathcal{V}_{\{2,3\}}\!=\!0.41\!\pm\!0.11$.
The experimental points adhere to the theoretical curve, being degraded slightly due to drifts in our control system.
\begin{figure}[ht]
\includegraphics[width=8cm]{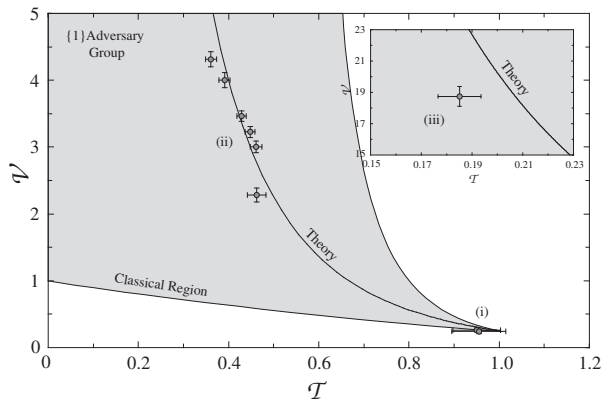}
\caption{ Experimental signal transfer $(\mathcal{T})$ and 
additional noise $(\mathcal{V})$ for the \{1\} adversary group, for increasing squeezing and additional Gaussian noise in the dealer protocol.
Solid line: theoretical curve with increasing squeezing or additional Gaussian noise. 
(i) Experimental point with no squeezing or additional noise, (ii) Experimental points with  squeezing of additional noise varied around $-4.5$~dB and $+3.5$~dB respectively (iii) and experimental point with $-4.5$~dB of squeezing and
$+18.6$~dB of additional noise with respect to the quantum noise limit. 
}\label{fig:fig5}
\end{figure}

Figure~\ref{fig:fig5} shows the $\mathcal{T}$ and $\mathcal{V}$ points for the adversary group \{1\} for increasing squeezing and additional Gaussian noise in the dealer protocol. 
Figure~\ref{fig:fig5} shows how the security of the scheme against individual players is enhanced by increasing either the squeezing or additional Gaussian noise. 
For no squeezing or additional Gaussian noise in the dealer protocol, the adversary group can obtain equal information about the secret state as the \{2,3\} authorized group with $\mathcal{T}_{\{1\}}\!=\!0.96\!\pm\!0.06$ and $\mathcal{V}_{\{1\}}\!=\!0.24\!\pm\!0.03$. 
The adversary group obtains almost no information about the secret state in the case of  $-4.5\!\pm\!0.2$~dB of squeezing and $18.6\!\pm\!3.8$~dB of additional noise in the dealer protocol, with $\mathcal{T}_{\{1\}}\!=\!0.19\!\pm\!0.01$ and $\mathcal{V}_{\{1\}}\!=\!18.7\!\pm\!0.6$.
To achieve this large amount of additional Gaussian noise, the optical parametric amplifiers are displaced with noise centered around the secret state frequency of 6.12~MHz. This demonstrates that the amount of information the adversary group obtains about the secret can be reduced to zero by increasing either the squeezing or additional Gaussian noise in the dealer protocol.  The spatial separation of the adversary group $\mathcal{T}$ and $\mathcal{V}$ points from that of the authorized group illustrates the 
information difference about the secret state obtained by both parties. 

\section{Conclusion}

In conclusion, we have demonstrated that for the $(2,3)$ 
threshold quantum state sharing scheme, there exists a 
class of ``disentangling" protocols which can be used to 
reconstruct the secret state. We have experimentally 
demonstrated that for this scheme, any two of the three players 
can form an authorized group to reconstruct the quantum state, 
achieving a fidelity averaged over all reconstruction permutations of 
$0.73 \pm 0.02$, a level achievable only using quantum resources.
We demonstrated that the entangled state and classical encoding 
techniques in the dealer protocol provide security against 
individual players, and in the case of finite squeezing in 
the dealer protocol, the security can be arbitrarily enhanced 
using classical encoding techniques. 

This demonstration of $(2,3)$ threshold quantum state 
sharing can be scaled up to $(k,n)$ threshold quantum
state sharing without increasing the number of active
elements (e.g. optical parametric amplifiers and
feed-forward devices)~\cite{Tyc03}. Although scaling up does
not result in an increase in the number of active
devices required by the players, the dealer does require
an increasing number of two-mode squeezed states to
hide the state being transmitted. Despite this challenge, 
extending beyond three players should be possible and
will elucidate the scaling properties of quantum state
sharing. Furthermore, it may be possible to substitute
one of the two squeezers in the dealer protocol
by an electro-optic feed-forward and an amplifier
as was done here for $(2,3)$ quantum state sharing, in
which case the two-squeezer requirement~\cite{Tyc03} could be 
relaxed.
This implementation of quantum 
state sharing broadens the scope of quantum information 
protocols, allowing the secure and robust transfer of quantum information 
and also
provides security against malicious parties or node and channel failures
in quantum information networks.  Teleported states, quantum computer output
states, and quantum keys used in quantum cryptography can all be
securely distributed using quantum state sharing.

The authors wish to thank Roman Schnabel, David Pulford and 
Hans Bachor for useful discussions and the support of the 
Australian Research Council, Australian Department of Defence and iCORE.

\end{document}